\documentclass[pra,aps,showpacs,twocolumn]{revtex4}

\usepackage{amssymb}
\usepackage{amsmath}
\usepackage{array}
\usepackage{graphicx}

\def\eps{\varepsilon}
\def\adj{^\dagger}
\def\ktau{(\vec{k},\tau)}
\def\komega{(\vec{k},i\omega{_n})}
\def\kaomega{(\vec{k},\omega)}

\renewcommand\vec[1]{{\mathbf #1}}
\def\mat[#1][#2]{{\hspace{.3ex}{\frac{}{}}^{#1}\underline{\underline{#2}}}}

\def\gmat[#1][#2]{\hspace{.3ex}{\frac{}{}}^{#1} #2}
\def\vv[#1][#2][#3]{\hspace{.3ex}{\frac{}{}}_{#1}^{#2} \underline{#3}}
\def\kink{e_{\vec{k}}}
\def\kinkq{e_{\vec{k}+\vec{q}}}

\def\Rb{${}^{87}\mathrm{Rb}$ }
\def\z{\textit{z}}
\def\mb{\mu_{\mathrm{B}}}
\def\TM{T_{\mathrm{M}}}
\def\Tc{T_{\mathrm{c}}}
\def\Bc{B_{\mathrm{c}}}
\def\TB{T_{\mathrm{BEC}}}
\def\kb{k_{\mathrm{B}}}
\def\nK{\;\mathrm{nK}}
\def\ve{\varepsilon}
\def\d{\mathrm{d}}

\begin{document}
\title{Static properties and spin dynamics of the
  ferromagnetic spin-1 Bose gas in magnetic field}
\author{Kriszti{\'a}n Kis-Szab{\'o}}
\affiliation{Department of Physics of Complex Systems, Roland E{\"o}tv{\"o}s
University, P{\'a}zm{\'a}ny P{\'e}ter s{\'e}t{\'a}ny 1/A, Budapest, H-1117}
\author{P{\'e}ter Sz{\'e}pfalusy}
\affiliation{Department of Physics of Complex Systems, Roland E{\"o}tv{\"o}s
University, P{\'a}zm{\'a}ny P{\'e}ter s{\'e}t{\'a}ny 1/A, Budapest, H-1117}
\affiliation{Research Institute for Solid State Physics and Optics of the
Hungarian Academy of Sciences, Budapest, P.O.Box 49, H-1525}
\author{Gergely Szirmai}
\affiliation{Research Group for Statistical Physics of the Hungarian Academy
  of Sciences, P{\'a}zm{\'a}ny P{\'e}ter S{\'e}t{\'a}ny 1/A, Budapest, H-1117}
\date\today
\begin{abstract}
  Properties of spin-1 Bose gases with ferromagnetic interaction in
  the presence of a nonzero magnetic field are studied. The equation
  of state and thermodynamic quantities are worked out with the help
  of a mean-field approximation. The phase diagram besides
  Bose--Einstein condensation contains a first order transition where
  two values of the magnetization coexist. The dynamics is
  investigated with the help of the Random Phase Approximation. The
  soft mode corresponding to the critical point of the magnetic phase
  transition is found to behave like in conventional theory.
\end{abstract}


\maketitle

\section{Introduction}

Atomic Bose gases consisting of atoms with spin 1 can exhibit a spinor
condensate, which has attracted considerable attention in recent years
\cite{SKea,Sea1,Sea2,Miesea,Stamper-Kurn2001a,Ho2,OM,LPB,HG1,HG2,HYip,
  HYin}. Such gases have interesting properties also before the
Bose--Einstein condensation (BEC) sets in. This is particularly true
when the coupling between the spin degrees of freedom prefers
ferromagnetic ordering. Such a system can be realized by the gas of
\Rb atoms. A \Rb atom has a nuclear spin of $j=3/2$ and an electron
spin of $s=1/2$, therefore its net spin can be $F=1$ or $F=2$. Since
the energy of the $F=1$ multiplet is smaller than that of the $F=2$,
the population in the $F=2$ multiplet dwindles at sufficiently low
temperatures resulting in a gas of purely spin-1 (3-component) bosons
\cite{Ho2,OM}.

Az zero external magnetic field it has been found that when lowering
the temperature a transition to ferromagnetic state can occur first
which can be first or second order as well depending on the strength
of the spin--spin interaction \cite{GK1}. The transition to the Bose
condensed phase can also be first or second order according to
mean-field theory results.  Note that the tendency towards
ferromagnetic ordering is present in the Bose gas formed by atoms with
nonzero spin even in the case when only a spin independent interaction
acts between the atoms (or the gas is ideal) \cite{Suto1,YL1,EL1,SC1}.
Effects of an external magnetic field has been studied for the free
Bose gas in Ref.  \cite{SC1}. In the present paper a gas of spin-1
bosons is investigated in the presence of an interaction of
ferromagnetic type and a nonzero magnetic field.

The system is supposed to be translationally invariant with a
homogeneous magnetic field pointing to the \z-directionand. The
Hamiltonian takes the following form:
\begin{multline}
\label{eq:ham}
  {\mathcal H}=\sum_{\genfrac{}{}{0pt}{2}{\vec{k}}{r,s}}
  \Big[(e_{\vec{k}}-\mu)\delta_{rs} -g \mb B\,
  (F_z)_{rs}\Big] a_r^\dagger(\vec{k})
  a_s(\vec{k})\\+\frac{1}{2V}\sum_{\genfrac{}{}{0pt}{2}{\vec{k}_1+
    \vec{k}_2=\vec{k}_3+\vec{k}_4}{r,s,r',s'}}a^\dagger_{r'}(\vec{k}_1)
  a^\dagger_r(\vec{k}_2)V^{r's'}_{rs}a_s(\vec{k}_3)a_{s'}
  (\vec{k}_4),
\end{multline}
where $a_r\adj(\vec{k})$ and $a_r(\vec{k})$ create and destroy
one-particle plane wave states with momentum $\vec{k}$ and spin
projection $r$. The spin index $r$ refers to the eigenvalue of the
\z-component of the spin operator and can take values from $+,0,-$. In
this basis the spin operators are given by:
\begin{multline}
  F_x=\frac{1}{\sqrt{2}}\left [ \begin{array}{c c c}
      0&1&0\\
      1&0&1\\
      0&1&0
    \end{array}\right ], \quad
  F_y=\frac{1}{\sqrt{2}}\left [ \begin{array}{c c c}
      0&-i&0\\
      i&0&-i\\
      0&i&0
    \end{array}\right ],\\
    F_z=\left [ \begin{array}{c c c}
      1&0&0\\
      0&0&0\\
      0&0&-1
    \end{array}\right ].
\end{multline}
In Eq. \eqref{eq:ham} $\kink=\hslash^2k^2/2M$ refers to the kinetic
energy of an atom ($M$ is the mass of an atom), $\mu$ to the chemical
potential, $g$ to the gyromagnetic ratio, $\mu_{\mathrm{B}}$ to the
Bohr magneton, B to the modulus of the homogeneous magnetic field, $V$
is the volume of the system and $V^{r's'}_{rs}$ the Fourier transform
of the two particle interaction potential, which for the low
temperature, dilute gas can be modeled by the momentum independent
s-wave scattering amplitude given for spin-1 bosons by \cite{BS1,SG}:
\begin{equation}
  \label{eq:pseudopot}
    V^{r's'}_{rs}=c_n\delta_{rs}\delta_{r's'}+c_s(\vec{F})_{rs}
    (\vec{F})_{r's'},
\end{equation}
with parameters:
\begin{subequations}
  \label{eqs:pspars}
  \begin{align}
    c_n&=\frac{4\pi\hslash^2}{M}\frac{a_0+2a_2}{3},\\
    c_s&=\frac{4\pi\hslash^2}{M}\frac{a_2-a_0}{3}.
  \end{align}
\end{subequations}
The parameters $a_0>0$ and $a_2>0$ are the scattering length in the
total hyperfine spin channel zero and two, respectively. Note, that
$c_s<0$ for the gas of \Rb atoms \cite{KBG}. For such a system it is
energetically favorable to align the spins along one direction, i.e.
the system has a ferromagnetic coupling \cite{Ho2,OM}. In this paper
we assume such a system.

The outline of the paper is as follows. In section \ref{sec:eqst}. the
Hartree equation of state of the dilute and low-temperature,
interacting spin-1 Bose gas is given in the presence of a nonzero
magnetic field. In section \ref{sec:magsusc}. the magnetic
susceptibility of the system is investigated in the uncondensed phase.
Section \ref{sec:cfun} is devoted to the determination of correlation
functions of generalized density operators, which are used in section
\ref{sec:wndsusc} to express the linear response functions of the
system. With the help of the response functions further static
properties and the spin dynamics of the system are also investigated.
In Section \ref{sec:sum} the results are summarized.

\section{Equation of state}
\label{sec:eqst}

For the spin-1 Bose gas the Hartree approximation \cite{SzSz2} yields
a plausible set of equations, which form the equation of state of the
system. The first equation in the set expresses the total density as
the sum of the density of the condensate and that of the different
spin projections of the non-condensate and reads as:
\begin{equation}
  \label{eq:eqstat1}
  n=n_0+n'_{+}+n'_{0}+n'_{-},
\end{equation}
where $n=N/V$ the total density of atoms in the gas, $n_0=N_0/V$ the
density of the Bose--Einstein condensed atoms, and
\begin{subequations}
  \label{eqs:hartreeloop}
\begin{equation}
  n'_{r}=\sum_{\vec{k}} n'_{\vec{k},r},
\end{equation}
is the density of the noncondensed atoms with spin projection $r$,
where
\begin{align}
  n'_{\vec{k},r}&=\frac{1}{e^{\beta \ve^{\mathrm{H}}_{\vec{k},r}}-1},\\
  \ve^{\mathrm{H}}_{\vec{k},r}&=\kink-\mu+c_n n+r(m c_s-g\mb B).
  \label{eq:hartreeen}
\end{align}
\end{subequations}
In the last equation $m$ is the magnetization density of the system
[see below] and $\beta=1/\kb T$ is the inverse temperature. The term
$r m c_s$ can be interpreted as the energy shift of the internal state
$r$ arising from a ``molecular field''. The formula is quite plausible
and can also be derived from a self-consistent Hartree approximation
in the Green's function technique. The next equation of the set is the
expression of the magnetization density $m$, and it takes the
following form:
\begin{equation}
  \label{eq:eqstat2}
  m=n_0+n'_{+}-n'_{-},
\end{equation}
since the condensate spinor points to the \z-direction. In the case,
when $n_0=0$, i.e. the system is not Bose--Einstein condensed Eqs.
\eqref{eq:eqstat1} and \eqref{eq:eqstat2} together with the
expressions \eqref{eqs:hartreeloop} form a closed set of equations,
and in the knowledge of the temperature, magnetic field and particle
density can be solved to the chemical potential and the magnetization
density. However, with lowering the temperature, one arrives at a
point, when the lowest bound of the energy expression will be zero:
$\ve^{\mathrm{H}}_{0,+}=0$, i.e. the chemical potential will reach the
value of $\mu=c_n n +c_s m - g \mb B$, which means that
$n'_{\vec{k},+}=[\exp(\beta\kink)-1]^{-1}$, so the system undergoes
Bose--Einstein condensation. This remains true at lower temperatures,
i.e.
\begin{equation}
  \label{eq:eqstat3}
  0=n_0\big[-\mu+c_n n +c_s m - g \mb B\big]
\end{equation}
holds. The multiplicative factor $n_0$ is used to make the equation
valid for the high temperature phases as well, where $n_0=0$. In
conclusion Eqs. \eqref{eq:eqstat1}, \eqref{eq:eqstat2} and
\eqref{eq:eqstat3} together with the expressions
\eqref{eqs:hartreeloop} form a closed set of equations for all
possible temperature values and is considered as the equation of state
of the spin-1 Bose gas with ferromagnetic interaction in the presence
of a magnetic field. Our equation of state compares for zero magnetic
field with that obtained by Gu and Klemm \cite{GK1} in a completely
different approach, while for $c_s=0$ it becomes similar to that of
the free Bose gas treated in a nonzero magnetic field by Simkin and
Cohen \cite{SC1}.

For the solution of the equation of state let us restrict ourselves
for the fixed particle density, magnetic field and temperature case,
i.e.  Eqs. \eqref{eq:eqstat1}, \eqref{eq:eqstat2} and
\eqref{eq:eqstat3} are solved for $(\mu,m,n_0)$ with $(n,B,T)$ fixed.
It is important to note that with introducing $\mu'=\mu-c_n n$ the
effect of $c_n n$ can be incorporated to the chemical potential for
fixed particle density. Detailed investigations show that the magnetic
separation found is accompanied by the separation of density. The
density difference, however between the phases is smaller by a factor
of order $|c_s|/c_n$ than the density itself. To concentrate to the
magnetic properties alone we disregard the phase separation in the
density. The phase diagram including phase separation will be
published elsewhere \cite{SzKSz}.

%% (Note, however, that $c_n$ causes a density
%% dependent shift in the chemical potential, which can alter even
%% certain thermodynamic quantities. The similarity to the free Bose gas
%% when $c_s=0$, mentioned above, is valid only in this restricted
%% sense.)

\begin{figure*}[!t]
  \centering
  \includegraphics{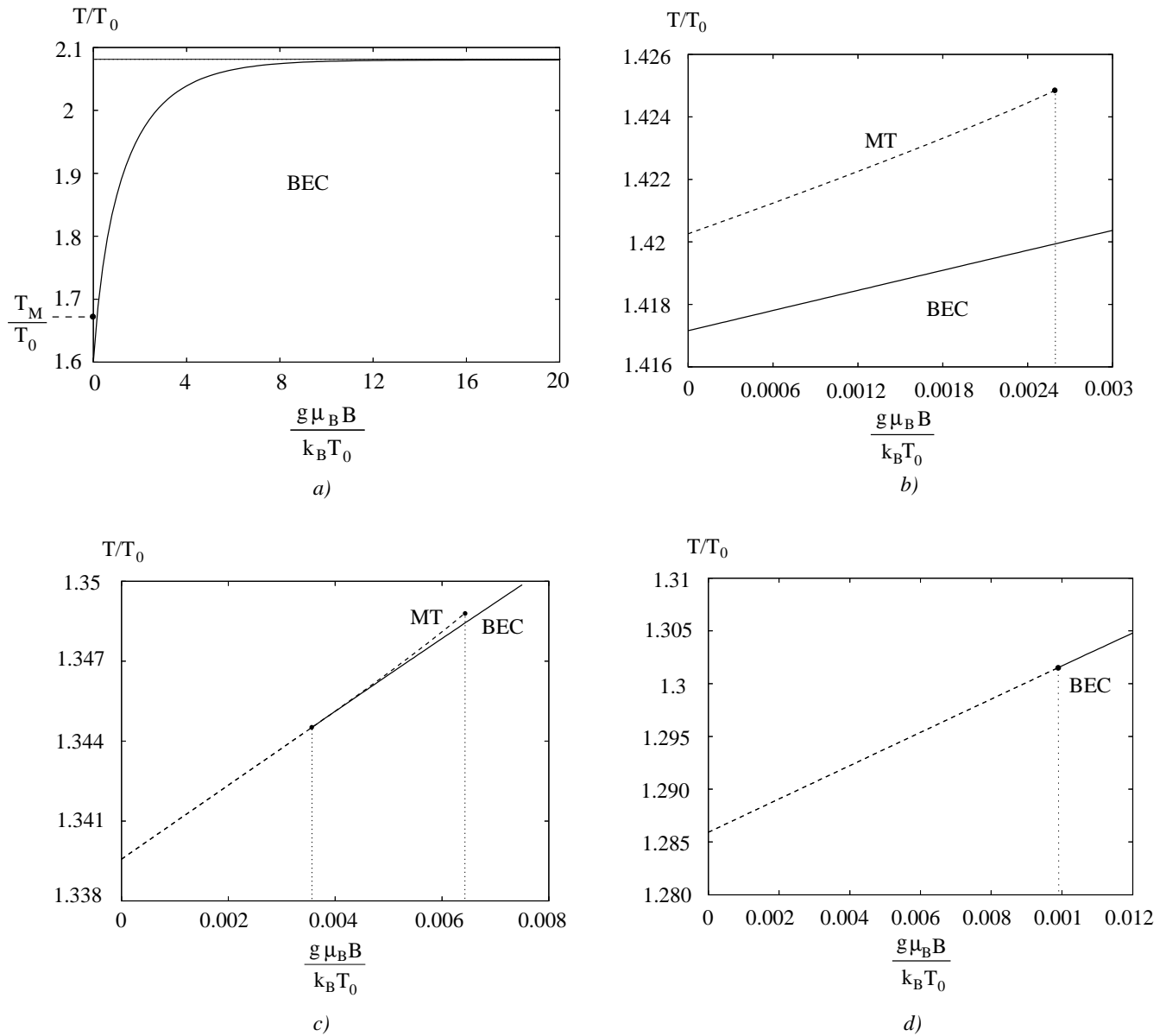}
  \caption{Possible phase diagrams of the system: a) for
    $\epsilon_s>\epsilon_s^{(1)}$, b) for $\epsilon_s^{(2)} <
    \epsilon_s < \epsilon_s^{(1)}$, c) for $\epsilon_s^{(3)} <
    \epsilon_s < \epsilon_s^{(2)}$, d) for
    $\epsilon_s\le\epsilon_s^{(3)}$. The dashed line symbolizes a
    first-order phase transition, while the continuous line refers to
    a continuous phase transition.}
  \label{fig:pde1}
\end{figure*}

The BEC transition temperature of the homogeneous spin-1 Bose gas is
given by
\begin{equation}
\label{eq:trtni}
  T_0=\frac{2\pi\hslash^2}{\kb M}\left[\frac{n}{3
      \zeta\big(\frac{3}{2}\big)}
  \right]^{\frac{2}{3}},
\end{equation}
and by taking $T_0$ to be, e.g. $200\nK$ (as is typical in experiments
\cite{BSC,HGT}) the particle density is given by fundamental constants
(apart from the mass of the atoms in the gas). For the case of \Rb
atoms, with $T_0=200\nK$ the particle density by Eq. \eqref{eq:trtni}
is $n=10^{20}\;\mathrm{m^{-3}}$. It is convenient to introduce the
following dimensionless parameter:
\begin{equation}
  \label{eq:epss}
  \epsilon_s=\frac{n |c_s|}{\kb T_0},
\end{equation}
which can be interpreted as a sort of mean-field energy in units of
$\kb T_0$.

By solving the equation of state the domain of $\epsilon_s$ can be
divided into four parts according to the character of the occurring
magnetic and BEC phase transitions. In the first region, when
$\epsilon_s$ is greater than a certain value
($\epsilon_s>\epsilon_s^{(1)}\approx1.22$) there exist a
paramagnetic-ferromagnetic transition for $B=0$ at $T=\TM$ above the
BEC transition $\TB$. Such a phase diagram is depicted in Fig.
\ref{fig:pde1} a) for $\epsilon_s=1.4$. Both transitions are
continuous for these values of $\epsilon_s$; the magnetic transition
exist only in the absence of magnetic field and its critical
temperature is higher than the BEC transition temperature \cite{GK1}.

If the value of $\epsilon_s$ is smaller than $\epsilon_s^{(1)}$ the
magnetic transition at $B=0$ becomes a first order one. Therefore the
point $(\epsilon_s=\epsilon_s^{(1)},B=0)$ can be understood as a
tricritical point for the magnetic transition, since it separates the
region of first-order and continuous transitions. The first order
transition survives even in the presence of a small magnetic field
leading to the coexistence of two phases with different
magnetizations.  The transition, called the magnetic transition (MT)
in the following, ends in a magnetic critical point (MCP), specified
by the critical value of the temperature $\Tc$ and that of the
magnetic field $\Bc$ for fixed $\epsilon_s$.  Note that a first order
phase transition due to an external field coupled to the order
parameter was observed in the ferroelectric
$\mathrm{Ba}\,\mathrm{Ti}\,\mathrm{O}_3$ already in
1953\cite{Merz1,Devonshire1}. If at the same time
$\epsilon_s>\epsilon_s^{(2)}\approx0.78$, BEC remains a continuous
phase transition and the phase diagram of this region looks like
plotted in Fig. \ref{fig:pde1} b) for $\epsilon_s=0.9$.

\begin{figure}[t!]
  \centering
  \includegraphics{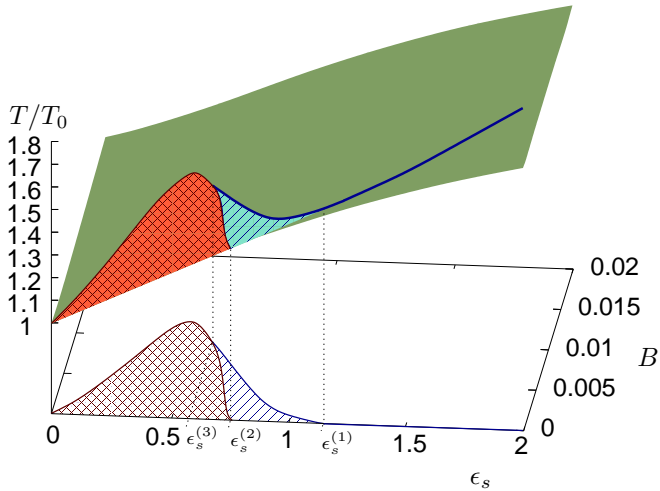}
  \caption{The phase diagram of the spin-1 Bose gas in the space of
    $(\epsilon_s,B,T)$. The temperature is made dimensionless with the
    help of $T_0$, while the magnetic field is given in units of $\kb
    T_0/ g \mb$. The solid surface indicates a continuous BEC
    transition, while in the cross-hatched region BEC is of first
    order. The simply hatched region, outside the region of first
    order BEC, refers to the first order magnetic transition and is
    lifted slightly from the surface of continuous BEC. At the heavy
    curve the magnetic transition is continuous, and its transition
    temperature is above that of the BEC. The surfaces of first order
    transitions and the curve of the continuous magnetic transition is
    projected to the $(\epsilon_s,B)$ plane for enlightenment. The
    relevant $\epsilon_s$ values are also indicated [compare with
    Figs.  \ref{fig:pde1} a)-d)].}
  \label{fig:pde5}
\end{figure}

If $\epsilon_s$ becomes smaller than $\epsilon_s^{(2)}$, BEC becomes
also of first order at $B=0$. However for $g\mb B/\kb T_0\gg 1$ the
system behaves like the gas of noninteracting scalar particles with a
continuous BEC phase transition at $3^{2/3}T_0$, therefore there must
be a critical magnetic field value, $B_c^{(1)}$, which is also a
tricritical point, but in the sense of the BEC transition, which
separates the first order BEC in small magnetic field from the
continuous BEC in strong magnetic field.  Such a phase diagram is
plotted in Fig.  \ref{fig:pde1} c) for $\epsilon_s=0.72$. As long as
$\epsilon_s > \epsilon_s^{(3)}\approx0.62$ there exist another
critical magnetic field value ($B_c^{(2)}$)above which there is no
magnetic transition (magnetization behaves analytically).

For $\epsilon_s\le\epsilon_s^{(3)}$ the two critical magnetic field
values are equal ($B_c^{(1)}=B_c^{(2)}$). A phase diagram for this
situation is shown in Fig. \ref{fig:pde1} d) for $\epsilon_s=0.6$.

After discussing the above four possible situations (according to the
value of $\epsilon_s$), it is worthwhile to plot the full phase
diagram of the system, where $\epsilon_s$ is also considered as a
variable besides $T$ and $B$. It can be seen in Fig.  \ref{fig:pde5}.
The above four figures can be obtained as sections of Fig.
\ref{fig:pde5} made with a fixed $\epsilon_s$. It is important to
note that inside the simply-hatched region on the $(\epsilon_s,B)$
plane of Fig. \ref{fig:pde5} the magnetic transition is of first order
and its transition temperature is higher than that of the continuous
BEC (of the same $\epsilon_s$ and $B$ value).

\section{Magnetic susceptibility in the uncondensed phase}
\label{sec:magsusc}

With differentiating the equation of state \eqref{eq:eqstat1} and
\eqref{eq:eqstat2} with respect to the thermodynamic quantities $T$,
$B$, $n$ one can obtain generalized susceptibilities of the system,
such as $(\partial n/\partial \mu)_{T,B}$ or $(\partial m/\partial
B)_{T,n}$, etc. (the subscripts are referring to variables kept
constant during differentiation). Focusing on the magnetic transition
here we restrict ourselves to the magnetic susceptibility of the
system with fixed particle number, i.e. $(\partial m/\partial
B)_{T,n}$. The region of interest in the $(\epsilon_s, B)$ plane is
also restricted to the simply hatched region on Fig. \ref{fig:pde5}
and to the heavy line with $(\epsilon_s>\epsilon_s^{(1)},B=0)$, or
equivalently to the situations depicted on Figs.  \ref{fig:pde1}
a)-c), where one can find a purely magnetic transition above the BEC
transition temperature. The temperature is also restricted above
$\TB(\epsilon_s,B)$.

For this purpose let us first cast the expression of the density of
the noncondensed particles \eqref{eqs:hartreeloop} to a more explicit
form by performing the momentum integration:
\begin{equation}
  \label{eq:hartree2}
    n'_{r}=\frac{\Gamma\big(\genfrac{}{}{}{}{3}{2}\big)}{(2\pi)^2
      \lambda^3} F\big(\genfrac{}{}{}{1}{3}{2},\beta[-\mu+c_n n +
    r(m c_s-g\mb B)]\big),
\end{equation}
where $\lambda=\hslash/\sqrt{2 M \kb T}$ is the thermal wavelength of
the atom, $\Gamma(s)$ the gamma-function and $F(s,\gamma)$ is the
standard Bose--Einstein integral with parameter $s$ and argument
$\gamma$ \cite{Robinson}. The derivative of \eqref{eq:hartree2} with
respect to $B$ can be easily evaluated with the result:
\begin{multline}
 \label{eq:hartder}
  \bigg(\frac{\partial n'_{r}}{\partial B}\bigg)_{T,n}=\frac{\beta
    \Gamma\big(\genfrac{}{}{}{}{3}{2}\big)}{(2\pi)^2\lambda^3}
  F\big(\genfrac{}{}{}{1}{1}{2},\beta[-\mu'+r(m c_s-g\mb B)]\big)\\
   \times \Bigg[\bigg(\frac{\partial \mu}{\partial B}\bigg)_{T,n}
    + r\;\bigg(g \mb-c_s\bigg(\frac{\partial m}{\partial B}\bigg)_{T,n}
    \bigg)\Bigg],
\end{multline}
where $\mu'=\mu-c_n n$ was introduced for simpler notation, and the
relation $\d F(s,x)/\d x=-F(s-1,x)$ was used
\cite{Robinson}. Differentiating Eqs. \eqref{eq:eqstat1} and
\eqref{eq:eqstat2} with respect to $B$ with $n$ and $T$ held fixed one
arrives at a system of two equations for the quantities $(\partial
m/\partial B)_{T,n}$ and $(\partial \mu/\partial B)_{T,n}$, from which
the former one can be expressed as
\begin{equation}
  \label{eq:susc}
  \bigg(\frac{\partial m}{\partial B}\bigg)_{T,n}=g\mb\frac{(P+R)\;Q+
  4P\;R}{P+Q+R+c_s\big[(P+R)\;Q+4P\;R\big]},
\end{equation}
with
\begin{subequations}
  \label{eq:PQR}
  \begin{align}
    P&=\frac{\beta\Gamma\big(\genfrac{}{}{}{}{3}{2}\big)}
    {(2\pi)^2\lambda^3}F\big(\genfrac{}{}{}{1}{1}{2},
    \beta[-\mu' + (m c_s-g\mb B)]\big),\\
    Q&=\frac{\beta\Gamma\big(\genfrac{}{}{}{}{3}{2}\big)}
    {(2\pi)^2\lambda^3}F\big(\genfrac{}{}{}{1}{1}{2},
    -\beta\mu'\big),\\
    R&=\frac{\beta\Gamma\big(\genfrac{}{}{}{}{3}{2}\big)}
    {(2\pi)^2\lambda^3}F\big(\genfrac{}{}{}{1}{1}{2},
    \beta[-\mu' - (m c_s-g\mb B)]\big).
  \end{align}
\end{subequations}
The continuous magnetic transition is signalled by the divergence of
the susceptibility \eqref{eq:susc}, or equivalently by the vanishing
of its denominator:
\begin{equation}
  \label{eq:susdenvan}
  P+Q+R+c_s\big[(P+R)\;Q+4P\;R\big]=0,
\end{equation}
which is fulfilled along the thick curve of Fig. \ref{fig:pde5}. The
magnetic susceptibility shows interesting properties also near the
BEC. According to the expression \eqref{eq:susc} $(\partial m /
\partial B)_{T,n}$ develops a cusp at $\TB(B)$ similarly to the
susceptibility of the free Bose gas \cite{SC1} at fixed number of
particles.

\section{Correlation functions of generalized density operators}
\label{sec:cfun}

Consider the following density operators
\begin{subequations}
  \label{eqs:densops}
  \begin{align}
    n(\vec{k})&=\sum_{\vec{q},r} a\adj_r(\vec{k}+\vec{q})
    a_r(\vec{q}),\\
    \mathcal{F}_z(\vec{k})&=\sum_{\vec{q},r,s}(F_z)_{r,s}
    a\adj_r(\vec{k}+\vec{q})a_s(\vec{q}).
  \end{align}
\end{subequations}
The former one is the particle density operator, while the latter one
is the longitudinal magnetization density operator. Other generalized
density operators can be defined as well for the spin-1 Bose gas, as
done e.g. in Ref.  \cite{SzSz2}, but for the purposes of this paper
the above two is enough. The correlation functions of the generalized
density operators \eqref{eqs:densops} are defined (for $\vec{k}\neq0$)
as:
\begin{subequations}
  \label{eqs:dc}
  \begin{align}
    &D_{nn}\ktau=-\Big< T_\tau\big[ n\ktau n\adj(\vec{k}, 0)
    \big]\Big>,\label{eq:d1}\\
    &D_{zz}\ktau=-\Big< T_\tau\big[ {\mathcal{F}}_z\ktau
    {\mathcal{F}}_z\adj(\vec{k}, 0)\big]\Big>,\label{eq:d2}\\
    &D_{nz}\ktau=-\Big< T_\tau\big[ n\ktau{\mathcal
      {F}}_z\adj(\vec{k}, 0)\big]\Big>,\label{eq:d3}
  \end{align}
\end{subequations}
with $\tau$ being the imaginary time and $T_\tau$ the $\tau$ ordering
operator (see e.g. Ref. \cite{FW}). After the usual Fourier
transformation into the Matsubara frequency representation these
correlation functions in general satisfy the following equations
\cite{SzSz2}:
\begin{subequations}
  \label{eqs:D0eqk}
  \begin{align}
    \label{eq:Dnneq1}
    D_{nn}&=\hslash\Pi_{nn}+c_n\Pi_{nn}D_{nn}+c_s\Pi_{nz}D_{nz},\\
    \label{eq:Dzzeq1}
    D_{zz}&=\hslash\Pi_{zz}+c_n\Pi_{nz}D_{nz}+c_s\Pi_{zz}D_{zz},\\
    \label{eq:Dnzeq1}
    D_{nz}&=\hslash\Pi_{nz}+c_n\Pi_{nn}D_{nz}+c_s\Pi_{nz}D_{zz},
  \end{align}
\end{subequations}
with the polarization functions:
\begin{subequations}
  \label{eqs:polparts}
  \begin{align}
    \Pi_{nn}\komega&=\sum_{r,s}\Pi^{rr}_{ss}\komega,\\
    \Pi_{zz}\komega&=\sum_{r,s}r\;s\;\Pi^{rr}_{ss}\komega,\\
    \Pi_{nz}\komega&=\sum_{r,s}s\;\Pi^{rr}_{ss}\komega.
  \end{align}
\end{subequations}
The polarization part $\Pi^{rr}_{ss}\komega$ is the contribution of
interaction line irreducible Feynman diagrams, which can connect to an
interaction line $V^{a,b}_{r,r}$ from the right and to an interaction
line $V^{s,s}_{c,d}$ ($a,b,c,d$ arbitrary spin indices) from the left
\cite{SzSz2}. The interaction with four indices is defined in Eq.
\eqref{eq:pseudopot}. (All correlation functions and polarization
functions in Eqs.  \eqref{eqs:D0eqk} depend on the $\komega$
variables, which are omitted for the sake of brevity.) The solution of
Eqs.  \eqref{eqs:D0eqk} reads as:
\begin{subequations}
  \label{eqs:dsol}
  \begin{align}
    D_{nn}&=\hslash\frac{\Pi_{nn}(1-c_s\Pi_{zz})+c_s\Pi_{nz}^2}
    {\det\mat[0][\eps]},\label{eq:dsol1}\\
    D_{nz}&=\hslash\frac{\Pi_{nz}}{\det\mat[0][\eps]},
    \label{eq:dsol2}\\
    D_{zz}&=\hslash\frac{\Pi_{zz}(1-c_n\Pi_{nn})+c_n\Pi_{nz}^2}
    {\det\mat[0][\eps]}\label{eq:dsol3},
  \end{align}
  with the abbreviating notation
\begin{equation}
  \det\mat[0][\eps]=\left(1-c_n\Pi_{nn}\right)
  \left(1-c_s\Pi_{zz}\right)-c_nc_s\Pi_{nz}^2.
\end{equation}
\end{subequations}

The expressions \eqref{eqs:dsol} are quite general. They are valid
both in the condensed and uncondensed phases and are contain no
approximations. Of course the choice of the actual form of the
polarization functions $\Pi^{rr}_{ss}$ selects between the different
approximations and possible phases.
%% The polarization functions consists of two parts in general:
%% \begin{equation}
%%   \label{eq:polsplit}
%%   \Pi^{rr}_{ss}\komega=\Pi^{(r)rr}_{ss}\komega+\Pi^{(s)rr}_{ss}\komega,
%% \end{equation}
%% where the first part is the regular part, which is also one-particle
%% line irreducible (apart from being interaction line irreducible),
%% while the so-called singular part, the latter term in Eq.
%% \eqref{eq:polsplit}, is one-particle line reducible (yet it is also
%% interaction line irreducible). This latter term arises only in the
%% Bose condensed phase \cite{SzSz2}, while in the uncondensed phase its
%% value is zero.
The correlation functions are evaluated in the framework of the Random
Phase Approximation (RPA) and for the uncondensed phase ($n_0=0$),
where the polarization function is taken as the contribution of the
bubble graph \cite{SzSz2}, which (for $\vec{k}$ and $\omega_n$ not
simultaneously zero) reads as:
%% \begin{subequations}
%%   \label{eqs:bubbles}
  \begin{equation}
    \label{eq:bubble}
    \Pi^{rr}_{ss}\komega=-\frac{\delta_{r,s}}{\hslash}\int
    \frac{\d^3 q}{(2\pi)^3}\frac{n'_{\vec{k}+\vec{q},r}-n'_{\vec{q},r}}
    {i\omega_n-\hslash^{-1}(\kinkq-\kink)}.
  \end{equation}
%% while the singular part (for $r=s=+$) takes the following form:
%% \begin{equation}
%%   \label{eq:singpol}
%%   \Pi^{(s)++}_{++}\komega=\frac{n_0}{\hslash}\frac{1}{(i\omega_n)^2-\hslash^{-2}\kink^2},
%% \end{equation}
%% \end{subequations}
%% and zero for $r\neq+$ or $s\neq+$.
The expressions for finite magnetic field are quite similar to those
of zero magnetic field \cite{SzSz2}.  The effect of the magnetic
field appears explicitly only in the contribution of the bubble
graph \eqref{eq:bubble} and only through $n'_{\vec{q},r}$, see Eq.
\eqref{eqs:hartreeloop}.

\section{Properties of wavenumber dependent susceptibilities}
\label{sec:wndsusc}

The dynamics of the system can be studied with the help of elementary
excitations. Spin dynamics is related to longitudinal and transverse
spin excitations. However at the magnetic transition, where the
susceptibility $(\partial m/\partial B)_{T,n}$ diverges longitudinal
spin dynamics plays the relevant role as the soft mode of the
transition. The spectrum of elementary excitations is related to
linear response functions of the system (see e.g. ref. \cite{FW}),
which can be obtained from the correlation functions by analytical
continuation in frequency through the real axis from above. From now
on we shall deal with such retarded correlation functions (response
functions). With the help of linear response theory the density and
magnetization fluctuations of the system are related to the external
potentials $\delta\mu\kaomega$ and $\delta B\kaomega$ by
\begin{subequations}
  \label{eqs:linresp}
  \begin{flalign}
    \lefteqn{\delta n\kaomega = -\hslash^{-1} D_{nn}\kaomega
      \delta\mu\kaomega}\nonumber \\&&-g \mb \hslash^{-1} D_{nz}
    \kaomega \delta B\kaomega,\label{eq:linresp1}\\
    \lefteqn{\delta m\kaomega = -\hslash^{-1}D_{zn}\kaomega
      \delta\mu\kaomega}\nonumber \\
    &&-g \mb \hslash^{-1} D_{zz}\kaomega \delta B\kaomega.
    \label{eq:linresp2}
  \end{flalign}
\end{subequations}

From Eq. \eqref{eq:linresp1}:
\begin{equation}
  \label{eq:lder1}
  \delta \mu \kaomega = -g\mb\frac{D_{nz} \kaomega}{D_{nn}\kaomega}
  \delta B\kaomega-\hslash\frac{\delta n\kaomega}{D_{nn}\kaomega}.
\end{equation}
Substituting it to Eq. \eqref{eq:linresp2} one obtains the equation
\begin{multline}
  \label{eq:lder2}
  -\frac{\hslash}{g\mb}\delta m\kaomega = -\frac{\hslash}{g\mb}
  \frac{D_{nz}\kaomega}{D_{nn}\kaomega}\delta n\kaomega\\
  +\frac{D_{nn}\kaomega D_{zz}\kaomega
    -D_{nz}\kaomega D_{zn}\kaomega}{D_{nn}\kaomega}\delta B\kaomega.
\end{multline}
On this basis one can define
\begin{multline}
  \label{eq:dzzn}
  \hslash^{-1}D_{zz}^{(n)}\kaomega\\:=\hslash^{-1}\frac{D_{nn}\kaomega
    D_{zz}\kaomega-D_{nz}\kaomega D_{zn}\kaomega}{D_{nn}\kaomega}\\
  =\hspace{-2.17pt}\frac{\Pi_{nn}\kaomega\Pi_{zz}\kaomega-\Pi_{nz}\kaomega\Pi_{zn}\kaomega}{\Pi_{nn}\kaomega[1-c_s\Pi_{zz}\kaomega]+c_s
  \Pi_{nz}\kaomega\Pi_{zn}\kaomega},
\end{multline}
where Eqs. \eqref{eqs:dsol} were used to arrive at the last equation.

The magnetic susceptibility
\eqref{eq:susc} can be expressed with the help of the response
functions in the static limit ($\omega=0$), since
\begin{subequations}
  \label{eqs:suscwresp}
  \begin{align}
    \bigg(\frac{\partial m}{\partial B}\bigg)_{T,n} =
    &\lim_{\vec{k}\rightarrow0} \lim_{\omega\rightarrow0}
    \frac{\delta m\kaomega}{\delta B\kaomega};\\
    \Big(\text{with}\quad &\lim_{\vec{k}\rightarrow0}
      \lim_{\omega\rightarrow0}\delta n\kaomega = 0\Big).
  \end{align}
\end{subequations}
In this limit the contribution to $\delta m$ comes from the second
term of the r.h.s of Eq. \eqref{eq:lder2}. According to the definition
\eqref{eq:dzzn} it means that
\begin{equation}
  \label{eq:suscsumrule}
  \bigg(\frac{\partial m}{\partial B}\bigg)_{T,n} = -
  \frac{g\mb}{\hslash}\lim_{\vec{k}
    \rightarrow 0}D_{zz}^{(n)}(\vec{k},0).
\end{equation}
In the uncondensed phase the long wavelength limit of the static
polarization functions take the following forms \cite{FRSzG,SzSz2}:
\begin{subequations}
  \label{eqs:bubstat}
  \begin{align}
    \Pi^{(r)++}_{++}(\vec{k}\rightarrow 0, 0)=-P,\\
    \Pi^{(r)00}_{00}(\vec{k}\rightarrow 0, 0)=-Q,\\
    \Pi^{(r)--}_{--}(\vec{k}\rightarrow 0, 0)=-R,
  \end{align}
  and therefore
\end{subequations}
$\Pi_{nn}(\vec{k}\rightarrow 0, 0)=-P-Q-R$,
$\Pi_{nz}(\vec{k}\rightarrow 0, 0)=-P+R$ and
$\Pi_{zz}(\vec{k}\rightarrow 0, 0)=-P-R$. One can verify that the sum
rule \eqref{eq:suscsumrule} is equivalent to Eq. \eqref{eq:susc}.

When the density fluctuations are small the second term on the r.h.s.
of Eq. \eqref{eq:lder2} will dominate for small $\vec{k},\omega$
values. (Note, that neglecting $\delta n$, then $\delta \mu$ and
$\delta B$ are related according to Eq.  \eqref{eq:linresp1}).
Therefore we concentrate on the expression \eqref{eq:dzzn} of
$D_{zz}^{(n)}$.

In the case when the momentum independent part of the Hartree energies
\eqref{eq:hartreeen} can be assumed to be small, i.e. $|-\mu'|\ll k_B
T$ and $|m c_s - g \mb B|\ll k_B T$, and the wavenumber is much
smaller than the thermal wavenumber $k\lambda\ll 1$. Furthermore
the frequency is also sufficiently small: $\beta\hslash\omega\ll k
\lambda$, the contribution of the bubble graph \eqref{eq:bubble} can
be approximated as \cite{SzK,SzSz2}
\begin{subequations}
\label{eqs:bubapp}
\begin{multline}
  \label{eq:bubapproxdin}
  \Pi^{rr}_{rr}(u,\Omega)=-\frac{\beta}{4\pi^2\lambda^3}\bigg[
  \frac{\sqrt{\pi}}{2}F\big(\genfrac{}{}{}{1}{1}{2},\sigma_r\big)
    -\frac{\pi}{2\sqrt{\sigma_r}} \\+ i\pi\frac{1}{\Omega u+
      i2\sqrt{\sigma_r}}+i\pi\frac{u^2}{3(\Omega u+i2
      \sqrt{\sigma_r})^3}\bigg],
\end{multline}
with
\begin{align}
  \sigma_r&=-\beta[\mu'+r(m c_s -g \mb B)],\\
  u&=k\lambda,\\
  \Omega&=\frac{\hslash\omega}{\kink}=\frac{\beta\hslash\omega}{u^2}.
  \label{eq:omdef}
\end{align}
\end{subequations}
With these newly introduced notations the validity of the limiting
form \eqref{eq:bubapproxdin} is equivalent to $\sigma_r\ll 1$,
$|\Omega u|\ll 1$ and $|\Omega + i2\sqrt{\sigma_r}/u|\gg 1$. The
expression \eqref{eq:bubapproxdin} is to be used in the formulas
\eqref{eqs:polparts} and \eqref{eq:dzzn} to obtain an explicit form for the response function $D_{zz}^{(n)}(\vec{k},\omega)$.

In the static limit one can set $\Omega=0$ in the expression
\eqref{eq:bubapproxdin} safely with the result:
\begin{equation}
  \label{eq:bubapproxstat}
  \Pi^{rr}_{rr}(u,0)=-\frac{\beta}{4\pi^2\lambda^3}\bigg[
  \frac{\sqrt{\pi}}{2}F\big(\genfrac{}{}{}{1}{1}{2},\sigma_r\big)
    -\pi\frac{u^2}{24 \sigma_r^{3/2}}\bigg],
\end{equation}
which can be used in the polarization functions \eqref{eqs:polparts}
to obtain
\begin{align}
  \Pi_{nn}&=-\widetilde{P}-\widetilde{Q}-\widetilde{R};&
  \widetilde{P}&=P - \widetilde{p} u^2;&\widetilde{p}=
  \frac{\beta}{96 \pi \lambda^3}\frac{1}{\sigma_+^{3/2}},\nonumber\\
  \Pi_{nz}&=-\widetilde{P}+\widetilde{R};&
  \widetilde{Q}&=Q - \widetilde{q} u^2;&\widetilde{q}=
  \frac{\beta}{96 \pi \lambda^3}\frac{1}{\sigma_0^{3/2}},\nonumber\\
  \Pi_{zz}&=-\widetilde{P}-\widetilde{R};&
  \widetilde{R}&=R - \widetilde{r} u^2;&\widetilde{r}=
  \frac{\beta}{96 \pi \lambda^3}\frac{1}{\sigma_-^{3/2}}\nonumber
\end{align}
up to the order of $u^2$. Therefore the long wavelength limit of the
static susceptibility at $T=\Tc$, where Eq. \eqref{eq:susdenvan} holds
takes the following form:
\begin{multline}
  \label{eq:suscstatlw}
  D_{zz}^{(n)}(k,\omega=0)=\frac{\hslash}{(k\lambda)^2}\\
  \times\frac{Q(P+R)+4 PR}{\widetilde{p}+\widetilde{q}+\widetilde{r}
    +c_s[\widetilde{q} (P+R)+Q (\widetilde{p}+\widetilde{r})+ 4(P
    \widetilde{r} + \widetilde{p}R)]}.
\end{multline}
From Eq. \eqref{eq:suscstatlw} one can see that the critical
exponent $\eta$, defined as $C(k)\propto k^{-2+\eta}$ at the critical
temperature for the correlation function $C$ of the order parameter, is
zero ($\eta=0$).

\begin{figure}[t!]
  \centering
  \includegraphics{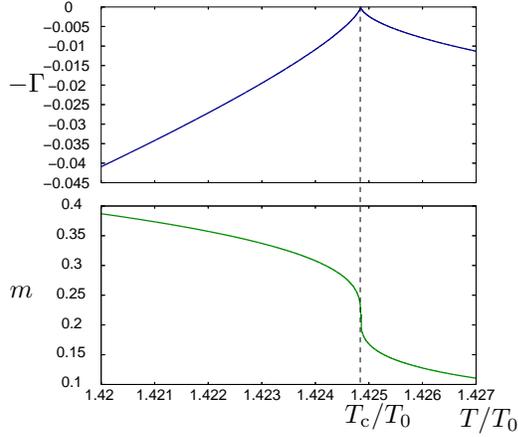}
  \caption{The damping rate $-\Gamma$ of the longitudinal spin 
    excitations (spin-density fluctuations) as a function of $T/T_0$
    at $B=\Bc$ and $\epsilon_s=0.9$ (upper panel). The damping rate is
    given in units of $\kb T_0/\hslash$. The magnetization density is
    plotted on the bottom panel for enlightenment.}
  \label{fig:longexc}
\end{figure}

To obtain the spectrum of the longitudinal spin fluctuations one has
to look for the poles of the response function \eqref{eq:dzzn}, or
equivalently look for the zeroes of its denominator. The resulting
equation is as follows:
\begin{equation}
  \label{eq:elexeq}
\Pi_{nn}\kaomega[1-c_s\Pi_{zz}\kaomega]+c_s
  \Pi_{nz}\kaomega\Pi_{zn}\kaomega = 0.  
\end{equation}
With the help of the numerical approximation of the analytic
continuation of the bubble graph \eqref{eq:bubble} \cite{SzK,SzSzK}
and with the help of Eqs.  \eqref{eqs:polparts}, Eq. \eqref{eq:elexeq}
can be solved directly. In the long wavelength limit one finds a
purely imaginary (overdamped) excitation spectrum depending linearly
on wavenumber:
\begin{equation}
  \label{eq:spect}
  \omega(k)=-i\Gamma\,k,
\end{equation}
with $\Gamma$ the velocity dimensional damping rate. The numeric
values of $\Gamma$ are given in Fig. \ref{fig:longexc} for
$\epsilon_s=0.9$, $B=B_c$ and temperature values above that of the
BEC.
One can clearly see the soft mode ($\Gamma=0$) at $B=\Bc$ and $T=\Tc$.

For analytical calculation at $\Bc$ and near $\Tc$ one can suppose
that $|\Omega u |\ll 2 \sqrt{\sigma_r}$ for all $r$, since the mode is
soft and neither of the $\sigma$-s is zero. In this case Eq.
\eqref{eq:bubapproxdin} can be further approximated. Taking only the
first three terms form the expression and casting the last one into
series of $\Omega u/2 \sqrt{\sigma_r}$ one arrives to:
\begin{equation}
  \label{eq:bubapproxsoft}
  \Pi^{rr}_{rr}(u,\Omega)=-\frac{\beta}{4\pi^2\lambda^3}\bigg[
  \frac{\sqrt{\pi}}{2}F\big(\genfrac{}{}{}{1}{1}{2},\sigma_r\big)
    +i\frac{\Omega u}{4\sigma_r}\bigg].
\end{equation}
%% The polarization functions \eqref{eqs:polparts} can be expressed by
%% introducing the following notations (similarly to those made in the
%% static limit):
%% \begin{align}
%%   \Pi_{nn}&=\!-\widehat{P}-\widehat{Q}-\widehat{R};&
%%   \widehat{P}&=P - i\widehat{p} \Omega u;&\widehat{p}=
%%   \frac{\beta}{16 \pi \lambda^3}\frac{1}{\sigma_+},\nonumber\\
%%   \Pi_{nz}&=-\widehat{P}+\widehat{R};&
%%   \widehat{Q}&=Q - i\widehat{q} \Omega u;&\widehat{q}=
%%   \frac{\beta}{16 \pi \lambda^3}\frac{1}{\sigma_0},\nonumber\\
%%   \Pi_{zz}&=-\widehat{P}-\widehat{R};&
%%   \widehat{R}&=R - i\widehat{r} \Omega u;&\widehat{r}=
%%   \frac{\beta}{16 \pi \lambda^3}\frac{1}{\sigma_-}.\nonumber
%% \end{align}
%% With their help Eq. \eqref{eq:elexeq} can be easily solved for
%% $\Omega$.
Using Eq. \eqref{eq:omdef} and \eqref{eq:spect} for $\Gamma$
one can obtain
\begin{multline}
  \label{eq:dampconst}
  \Gamma=\frac{\lambda}{\beta\hslash}\\
  \times\frac{P+Q+R+c_s[Q(P+R)+4PR]}{\widehat{p}+\widehat{q}+
    \widehat{r}+c_s[\widehat{q}(P+R)+Q(\widehat{p}+\widehat{r})+4(P\widehat{r}+\widehat{p}R)]},
\end{multline}
where $P$, $Q$, $R$ are given by Eqs. \eqref{eq:PQR} and
\begin{subequations}
  \label{eqs:whpqr}
  \begin{align}
    \widehat{p}&=\frac{\beta}{16 \pi \lambda^3}\frac{1}{\sigma_+},\\
    \widehat{q}&=\frac{\beta}{16 \pi \lambda^3}\frac{1}{\sigma_0},\\
    \widehat{r}&=\frac{\beta}{16 \pi \lambda^3}\frac{1}{\sigma_-}.
  \end{align}
\end{subequations}
The numerator of the damping parameter \eqref{eq:dampconst} is equal
to the denominator of the static, homogeneous susceptibility
\eqref{eq:susc}. Accordingly the damping rate, $\Gamma$, vanishes at
$T=\Tc$ (for $B=\Bc$, where the transition is continuous) and $\Gamma$
is proportional to the inverse of the susceptibility \eqref{eq:susc}
for $T\approx \Tc$, like in conventional theory.

\section{Conclusions and summary}
\label{sec:sum}

We have studied the statistical physics of the spin-1 Bose gas with
ferromagnetic coupling in the presence of a magnetic field both from
the static and dynamical points of view. The equation of state and the
phase diagram of the system was investigated both for the
Bose--Einstein condensed and uncondensed phases. The discussion of
other static quantities and the dynamics is concentrated to the
noncondensed phase, where the purely magnetic phase transition takes
place, which occurs prior the Bose--Einstein condensation if the
spin-flipping coupling constant ($|c_s|$) is large enough.

This magnetic transition was found to be of first order in general.
There is always a critical magnetic field value, where the magnetic
transition is continuous and above that there is no phase transition
at all (magnetization behaves analytically). There is also a
tricritical point (at $\epsilon_s=\epsilon_s^{(1)}\approx1.22$,
$T\approx1.58$), when the critical magnetic field reaches zero. Above
this value of $\epsilon_s$ the usual paramagnetic-ferromagnetic phase
transition occurs in zero external magnetic field. The static
susceptibility $(\partial m/\partial B)_{T,n}$ shows a divergence at
the critical points of the continuous MT, while it has only a cusp at
the point of the BEC.

The long wavelength dynamics also exhibits features of the MT. The
longitudinal spin excitation becomes a soft mode in accordance with
the diverging susceptibility in the vicinity of the critical point of
the MT. Moreover the soft mode behaves like in conventional theory;
close to $\Tc$ the damping rate is proportional to the inverse of the
susceptibility.

%\clearpage

\section{Acknowledgement}
\label{sec:ack}

We are grateful to L\'aszl\'o Sasv\'ari for a useful discussion and
for calling our attention to the relevant works in ferroelectric
systems. The present work has been supported by the Hungarian Research
National Foundation under Grant No. OTKA T046129.

\end{document}